\documentclass[apsrev4-1,prb,superscriptaddress,longbibliography,twocolumn]{revtex4-1}
\usepackage{colortbl}
\usepackage{color}
\usepackage{amsfonts,amsmath,amssymb}
\usepackage[dvipsnames]{xcolor}
\usepackage{tabularx,hhline,tikz,multirow,array}
\usepackage{bm,enumitem}
\usepackage{dcolumn,stmaryrd}
\usepackage[colorlinks=true,pdfborder={0 0 0},linkcolor=red,citecolor = blue]{hyperref}

\newcommand{\Sp}{{\bf S}}
\newcommand{\Tp}{{\bf s}}

\def\rv {{\bf r}}

\def \bea {\begin{eqnarray}}
\def \eea {\end{eqnarray}}

\begin{document}

\title{Skyrmion-skyrmion interaction induced by itinerant electrons in a ferromagnetic strip}

\author{E Iroulart}
\affiliation{Instituto de F\'isica de L\'iquidos y Sistemas Biol\'ogicos, CONICET, Facultad de Ciencias Exactas, Universidad Nacional de La Plata, 1900 La Plata, Argentina}
\affiliation{Departamento de F\'{i}sica, FCE, UNLP, La Plata, Argentina} 
\author{H. Diego Rosales}
\affiliation{Instituto de F\'isica de L\'iquidos y Sistemas Biol\'ogicos, CONICET, Facultad de Ciencias Exactas, Universidad Nacional de La Plata, 1900 La Plata, Argentina}
\affiliation{Departamento de F\'{i}sica, FCE, UNLP, La Plata, Argentina} 
\affiliation{Departamento de Ciencias B\'asicas, Facultad de Ingenier\'ia, UNLP, La Plata, Argentina}
\date{\today}

\begin{abstract}
Magnetic skyrmions are promising spin textures for building next-generation magnetic memories and spintronic devices. Nevertheless, one of the major challenges in realizing skyrmion-based devices is the stabilization of ordered arrays of these spin textures  in different geometries. Here we numerically study the skyrmion-skyrmion  interaction potential that arises due to the dynamics of itinerant electrons coupled to the magnetic texture in a ferromagnetic background with racetrack geometry. We consider different topological textures (ferromagnetic (FM) and antiferromagnetic (AFM)), namely: skyrmions, antiskyrmions and biskyrmions.
We show that at low electron filling, for sufficiently short separation, the skyrmions strongly couple each other yielding a bound-state bound by electronic dynamics. However, when the filling is increased, the interaction potential energy presents  local minima at specific values of the skyrmion-skyrmion distance. Each of these local minima corresponds to energetically stable positions of skyrmions which are ``protected'' by well-defined energy barriers. By inspecting the local charge density, we find that in the case of AFM skyrmions, the local antiferromagnetic nature prevents electronic penetration into the core, allowing the AFM skyrmions to be seen as infinite potential barriers for electrons.
\end{abstract}

\maketitle

\section{Introduction}

From the theoretical proposal in the last century and the subsequent experimental evidence in chiral magnets \cite{muhlbauer2009skyrmion,munzer2010skyrmion,yu2010real,yu2011near,heinze2011spon,seki2012observation,butenko2010stab}, magnetic skyrmions have garnered enormous interest due to their potential applications as promising information carriers in spintronics, including the design of racetrack memories\cite{fert2013skyrmions,zhang2015magnetic,zhang2020skyrmion,GUAN2022168852,Santos2021}. The advantages presented by skyrmions compared to other standard and extensively studied  magnetic textures (magnetic bubbles \cite{Eschenfelder1980}) are the small currents needed for their propagation, their smaller size ($\sim 1-100$ nm diameter) and their topological nature which provides a significant energy barrier to avoid skyrmion annihilation\cite{zhang2020skyrmion}.

\begin{figure}[htb]
 \centering
\includegraphics[width=0.9\columnwidth]{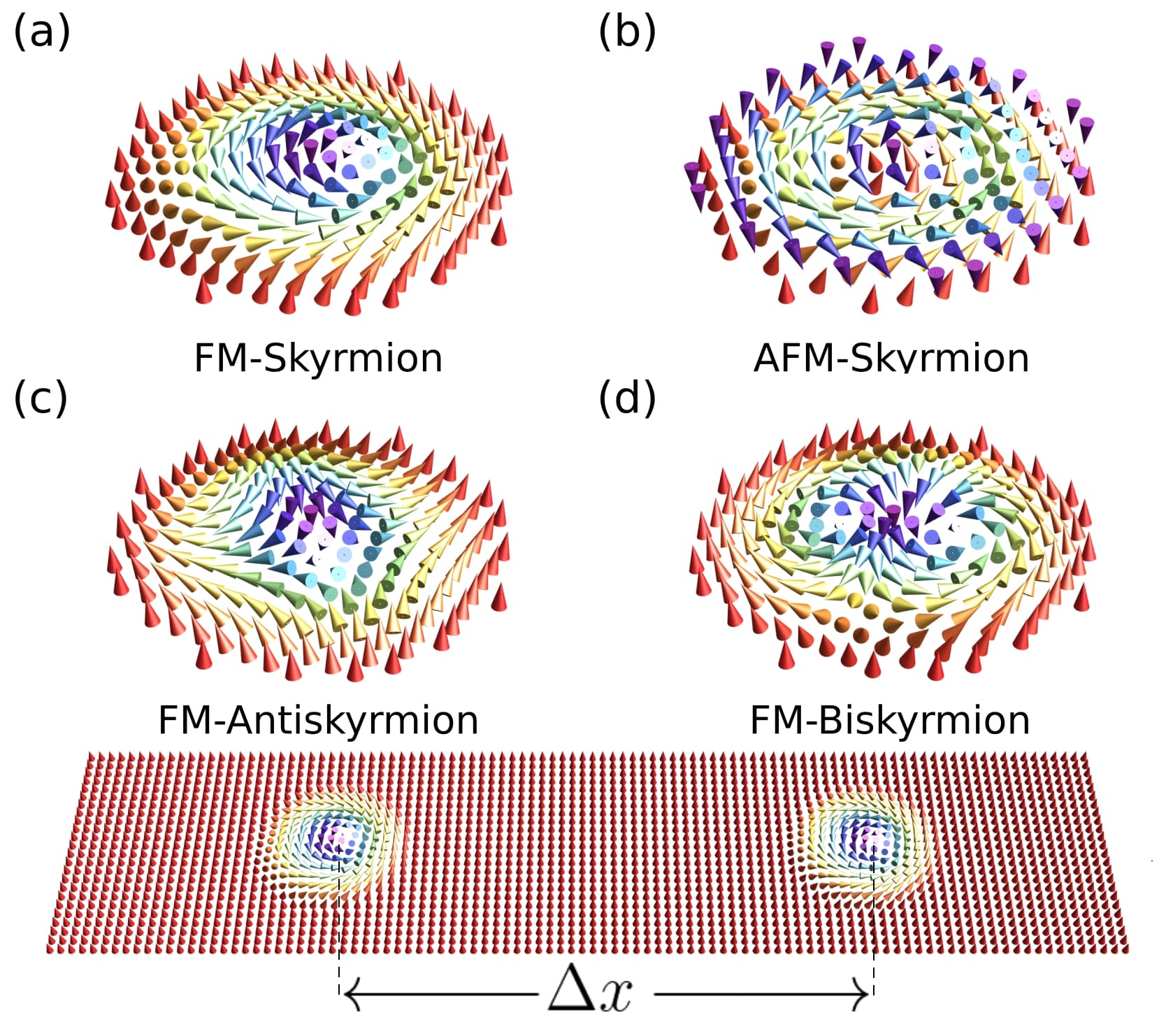}
\caption{(Color online) Schematics of various skyrmions; (a) ferromagnetic skyrmion, (b) antiferromagnetic skyrmion, (c) anti-skyrmion, and (d) biskyrmion. Bottom: skyrmions embedded in a collinear ferromagnetic background, separated at a distance $\Delta x$. between skyrmions centers} 
\label{fig:skyrmionTypes}
\end{figure}

Currently, there is a large family of topological skyrmion-like textures that includes the standard ferromagnetic skyrmions (FM) (Fig.\,\ref{fig:skyrmionTypes}a), antiferromagnetic skyrmions (AFM) \cite{rosales2015three,barker2016static,osorio2017composite,osorio2019skyrmions,villalba2019field,ShangGao2020Nat,Rosales2022a,legrand2020room,liu2021neel,mukherjee2021af} (Fig.\,\ref{fig:skyrmionTypes}b), antiskyrmions \cite{nayak2017magnetic} (Fig.\,\ref{fig:skyrmionTypes}c) and  biskyrmions \cite{yu2014bisk} (Fig.\,\ref{fig:skyrmionTypes}d), among others (for a recent review, see for example \cite{zhou2019magnetic,back2020road,gobel2020beyond}).

In particular, AFM skyrmions have  become the subject of intense focus in the context of antiferromagnetic spintronics \cite{jungwirth2016antiferromagnetic}. The interest in AFM skyrmions arises from the effect of coupling conduction electrons to the local magnetic background: the electrons accumulate a Berry phase as they travel through skyrmions spin configuration, which acts as a local effective magnetic field leading to topological Hall effect (THE) \cite{rosales2019frustrated,Djavid2020,tome2021topological}, spin Hall effect\cite{Gobel2017,Akosa2018} and the skyrmion Hall effect  \cite{barker2016static,chen2017skyrmion,zhang2016antiferromagnetic,jin2016dynamics}. Moreover, it has been shown that the spin-orbit coupling can be used to suppress the  skyrmion Hall effect\cite{Akosa2019}.
Compared with FM skyrmions, AFM ones can move along the direction of the driving force without showing the skyrmion Hall effect at velocities of several hundred meters per second \cite{gobel2018overcoming}. Therefore, they could be ideal information carriers for future spintronic devices, such as racetrack-type memories and logic computing devices.

Nevertheless, potential applications concerning skyrmions or the afore-mentioned analogues demand the understanding of the inter-skyrmion interactions and the effect of the space confinement in nano-sized geometries. In regard to this problem, there are a variety of methods focusing on the study of the interaction mediated by magnetic fluctuations. For example, the Thiele's approach predicts a short range repulsive force bewteen skyrmions \cite{Lin2013}. On the other hand, effective Ginzburg-Landau theory predicts a short range oscillating force \cite{Lin2016}. More recent studies have shown that magnetic frustration may induce the attractive interaction at short distances \cite{Rozsa2016} and inter-skyrmion interaction is always repulsive and decays exponentially at a large distance \cite{capic2020skyrmion,brearton2020magnetic}, while when skyrmions are nucleated in experiments they incidentally appear close to each other. 
The analysis of the contribution of the itinerant electrons on the skyrmion-skyrmion pair interaction has been previously studied in magnetic monolayers via the Ruderman-Kittel-Kasuya-Yosida (RKKY) exchange interaction \cite{bezvershenko2018stabilization} and in multilayer systems \cite{zhang2016magnetic,cacilhas2018coupling}, but the pairwise skyrmion-skyrmion interaction is yet to explore, raising the question of whether this electronic contribution can have drastic consequences in relation to the geometric/spacial confinement in racetrack geometries.

This paper seeks to address the study of the itinerant electrons-induced skyrmion-skyrmion interaction in a ferromagnetic background as a function of the skyrmion-skyrmion distance $\Delta x$ [see Fig.\,\ref{fig:skyrmionTypes}], and the filling $n$ (i.e. number of occupied states). Through numerical  calculations, we studied the influence of the skyrmion nature (both skyrmions FM or both AFM) and  its  toplogical charge $Q$: $Q=-1$ (skyrmion), $Q=+1$ (antiskyrmion) and $Q=-2$ (biskyrmion). The dependence of the skyrmion energy on the distance $\Delta x$  allows us to determine the stable positions of the skyrmions along the nanostripe. The special cases of the ferromagnetic (FM-FM) and antiferromagnetic (AFM-AFM) skyrmion-skyrmion interaction are discussed revealing a regular distribution of energy minima at specific values of separation indicating positional stability. We found that at low fillings both cases show  a very close behavior developing a very similar pattern of local minima; while at higher fillings, the AFM case presents a much more regular distribution of minima that indicates greater positional stability. In addition, we studied the local electronic charge distribution which allows us to highlight the fundamental difference between the FM and AFM cases: while in the first case, the electronic charge penetrates the core of the skyrmion as the filling increases, in the second case the antiferromagnetic nature of the magnetic texture prevents it, and thus the AFM skyrmions may be viewed as hard disks.  We found that in the AFM case, due to this characteristic the skyrmion-skyrmion problem presents a close relation with the double-well potential problem, showing quite similar energy vs $\Delta x$ curves.  Finally, we analyzed the effect of considering skyrmions of charge $Q\neq -1$.

\section{Model and Ans\"atze}
\label{sec:Model}
To set the stage, we consider a Kondo lattice model on square lattice of $L_x\times L_y$ sites with mixed boundary conditions, that is, periodic boundary condition along $x$ and open boundary condition along the $y$-axis.  The hopping term and the interaction of itinerant electrons with a magnetic texture are described by the following Hamiltonian,
\bea
\mathcal{H}=-t\sum_{\langle \rv,\rv'\rangle}\hat{c}^{\dagger}_{\rv,\sigma}\hat{c}_{\rv',\sigma}-J_H\sum_{\rv}\Sp_\rv\cdot\Tp_\rv ,
\label{eq:HtJ}
\eea

\noindent where the operator $\hat{c}^{\dagger}_{\rv,\sigma}$ ($\hat{c}_{\rv,\sigma}$) creates (annihilates) an electron with spin  $\sigma=\pm 1/2$  at site $\rv$. The first term $t$ is the transfer integral between nearest-neighbor sites of itinerant electrons. The second term describes the  Hund coupling between the spin of itinerant electrons  $\Tp_\rv=(1/2)\sum_{\sigma\sigma'}\hat{c}^{\dagger}_{\rv,\sigma}\vec{\sigma}_{\sigma\sigma'}\hat{c}_{\rv,\sigma'}$ and localized classical spins with coupling constant $J_H$, where $\vec{\sigma}$ is the vector of Pauli matrices.
In the computations, we take the lattice spacing $a=1$ and the hopping constant $t=1$ as an energy unit.

Hereafter, let's focus in the $J_H/t>>1$ regime, where the spin of the hopping electron is forced to align parallel to the local moment and the low-energy physics can be described by an effective Hamiltonian of spinless fermions as
\bea
\mathcal{H}_{eff}&=&-\sum_{\langle \rv,\rv'\rangle}t^{eff}_{\rv\rv'}\hat{d}^{\dagger}_{\rv}\hat{d}_{\rv'},
\label{eq:HtJeff}
\eea

\noindent where $\hat{d}^{\dagger}_{\rv}$ ($\hat{d}_{\rv}$) is the creation (annihilation) fermion operator and $t^{eff}_{\rv\rv'}$ is the effective transfer integral (see appendix for details).

For the purpose of studying the effective interaction between consecutive skyrmions on the racetrack geometry, we employ artificial skyrmion textures constructed using the finite size skyrmion ansatz (centered at the origin $\rv=(x,y)\equiv(0,0)$) \cite{osorio2019stability}: 

\bea
\text{Ferromagnetic:\quad}&&\Sp^{\text{FM}}=
\begin{bmatrix}
\sin(f)\cos(\phi)\\
\sin(f)\sin(\phi)\\
\cos(f) 
\end{bmatrix}\\
\label{eq:fm}
\text{Antiferromagnetic:\quad}&&\Sp^{\text{AFM}}=(-1)^{x+y}\Sp^{\text{FM}},
\label{eq:afm}
\eea
\noindent where $f\equiv f(r)=\pi(1-r/R)\Theta(R-r)$, $R$ is the skyrmion radius, $\Theta(R-r)$ is the Heaviside step function, $\phi\equiv \phi(\rv)=Q\times(\arctan(y/x)+\chi)$ with $\chi$ being the helicity and $Q$ the topological charge \cite{osorio2019stability,bogdanov1989thermodynamically}. With these ans\"atze it is possible to build different types of skyrmion configurations with a spacing $\Delta x$ between centers of two adjacent skyrmions. In the case of AFM skyrmions  (Eq.\,\ref{eq:afm}), we have used the most simple picture where the spin texture can be visualized as a superposition of two FM skyrmions coupled antiferromagnetically\cite{diaz2021majorana} (Fig.\,\ref{fig:skyrmionTypes}b). However, it should be mentioned that it is also possible to combine skyrmions on multiple sublattices \cite{rosales2015three,villalba2019field,ShangGao2020Nat,Rosales2022a,mukherjee2021af}.  

After diagonalizing  the electronic Hamiltonian in Eq.\~(\ref{eq:HtJeff}) by an unitary transformation $\mathcal{U}$, we compute the ground state $|GS\rangle=\prod_{\nu=1}^{n}|\nu\rangle$, where $\epsilon_{\nu}$ is the energy of the $\nu$-th electronic eigenstate, i.e. $\mathcal{H}_{eff}|\nu\rangle=\epsilon_{\nu}|\nu\rangle$. 

The ground state energy $E(\Delta x)$ and the on-site electron density $\rho_{\rv}$ at zero temperature for fixed separation $\Delta x$ (see top of  Fig. \ref{fig:EvsdxCurves}) at a fixed filling $n$, are given by $E(\Delta x)\equiv\langle GS|\mathcal{H}_{eff}|GS\rangle=\sum_{\nu=1}^{n}\epsilon_{\nu}$, $\rho_{\rv_i}=\langle GS|\hat{d}^{\dagger}_{\rv_i}\,\hat{d}_{\rv_i}|GS \rangle=\sum_{\nu=1}^{n}\mathcal{U}^{*}_{\nu,i}\mathcal{U}_{\nu,i}+\mathcal{U}^{*}_{\nu,i+N}\mathcal{U}_{\nu,i+N}$.

\section{Calculation of the interaction potential}
\label{se}

In this section, we present results for systems consisting of a pair of skyrmions with radius $R$ in a ferromagnetic background with size $L_x\times L_y$. Thorough the rest of this work, the term ``skyrmion'' is used to identify the $Q=-1$ topological texture, which can be either ferrromagnetic (FM) or antiferromagnetic (AFM).  We have focused on the case of a skyrmion-pair with the same topological charge $\{Q,Q\}=\{-1,-1\}$ (both FM-FM and AFM-AFM cases), $R=8$ and $\{L_x,L_y\}=\{300,25\}$. Furthermore, we have also analyzed the situation of different topological charges: $\{Q,Q\}=\{-1,+1\}$ and $\{Q,Q\}=\{-2,-2\}$; and larger systems sizes to rule out significant finite size effects. These results allow us to predict the  interaction potential between skyrmions assuming that they are displaced along the nanotrack as shown in Fig. \ref{fig:skyrmionTypes}.

\subsection{FM-FM and AFM-AFM skyrmion pair interactions with $Q=-1$}

\begin{figure}[htb]
\centering
\includegraphics[width=0.99\columnwidth]{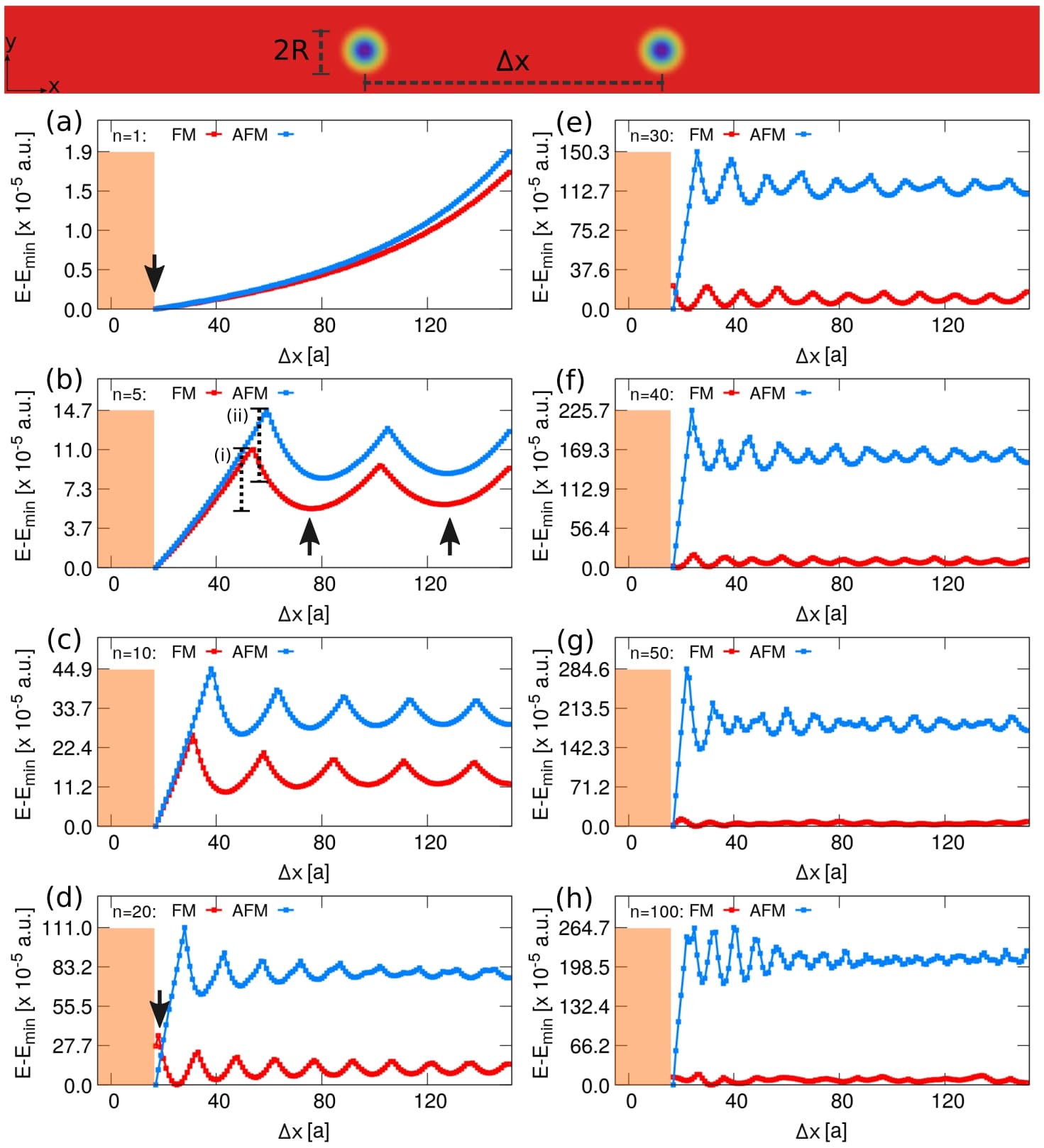}
\caption{(Color online) Top: An example of a skyrmion geometry performed to determine the skyrmion-skyrmion interaction potential. The length $\Delta x$ was progressively shortened to map out the energy increase of the system due to interactions between the skyrmions. (a)-(g) Skyrmion-skyrmion interaction potential as a function of the distance $\Delta x$ for different fillings: FM-FM (red) and AFM-AFM (blue).} 
\label{fig:EvsdxCurves}
\end{figure}

Firstly, we have considered the situation of a skyrmion-pair (FM-FM and AFM-AFM) with topological charge $Q=-1$ separated by a distance $\Delta x$. 
We have determined the interaction potential $E(\Delta x)$ as a function of the horizontal distance between the skyrmions at fixed filling $n$. Figures \ref{fig:EvsdxCurves} (a-h) depicts the behavior of $E(\Delta x)$ as a function of $\Delta x$ and for severals values of $n$. One can notice that at low fillings,  there is a minimum at $\Delta x=2\times R$ indicating that the more favorable arrangement corresponds to locating both SKs next to each other (black arrow in Fig. \ref{fig:EvsdxCurves}(a)). Therefore, for sufficiently short separation skyrmions strongly couple each other yielding a bound-state bound by electronic dynamics.
More remarkable is what happens  when $n$ is increased: a sequence of well-defined local minima  appear at specific values of $\Delta x$  (black arrows in Fig. \ref{fig:EvsdxCurves}(b)). Each of these local minima corresponds to energetically stable distances between skyrmions which are ``protected'' by well defined energy barriers (dashed lines (i) and (ii)).  This general scheme of local minima is observed in both cases (FM-FM and AFM-AFM pairs). Nevertheless, while in the FM case, for some fillings the interaction is repulsive at short distances (Fig. \ref{fig:EvsdxCurves}(d-e)), in the AFM case the global minimum always occurs in the bound-state configuration (with skyrmions next to each other). This would indicate that the bound state of AFM skyrmions would be much more stable against changes in electronic filling than in the FM case.

 \begin{figure}[htb]
 \centering
\includegraphics[width=0.99\columnwidth]{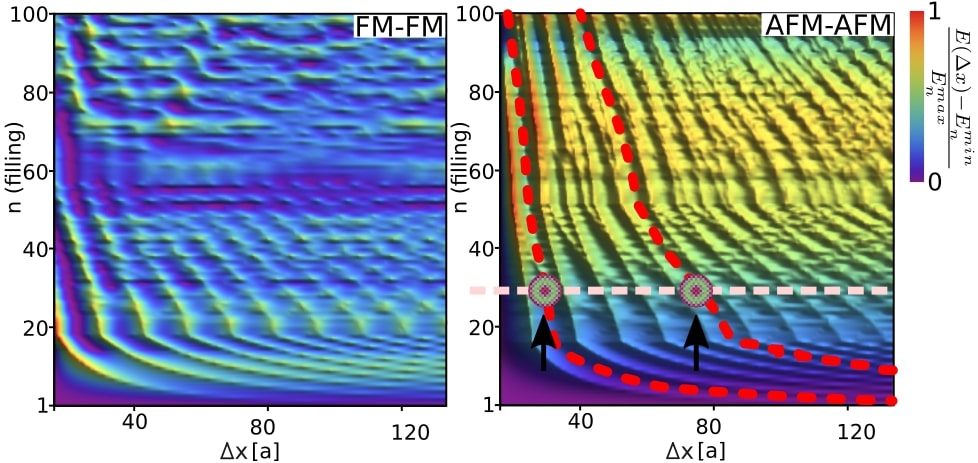}
\caption{(Color online) Phase diagram of the interaction potential $(E(\Delta x)-E^{min}_{n})/E^{max}_n$ as a function of the skyrmion separation $\Delta x$ and and the filling $n$ for ferromagnetic (left) and antiferromagnetic (right) skyrmions. On the right, we highlight two minima at specific values of $\Delta x$ and $n$. Dashed red lines are a guide to the eye. Arrows indicate a possible stable (in position) skyrmion configuration.}
\label{fig:EvsdxCurves2}
\end{figure}

To confirm this picture, we construct a density plot of the interaction potential with control parameters $\Delta x$ and $n$, for both cases (FM-FM and AFM-AFM, skyrmions), shown in Fig.~\ref{fig:EvsdxCurves2}. We plot the interaction potential $(E(\Delta x)-E^{min}_n)/E^{max}_n$ ($E^{min}_n$ and $E^{max}_n$ are the minimum  and maximum values of $E(\Delta x)$ that
takes a fixed $n$) where blue color indicates the lowest value $E(\Delta x)-E^{min}_n=0$. As it can be seen in Fig.~\ref{fig:EvsdxCurves2} (panels left and right), both the FM and AFM skyrmions develop well defined minima at low fillings. However, when filling $n$ is increased, in the case of FM skyrmions the local minima become irregular and diffuse (left), while in the AFM case the minima remain well defined forming a ribbed pattern as a function of $n$ for all cases (right).
This result strongly suggests that for AFM skyrmions the interaction with electrons generates an array of energetically stable separations of skyrmions along the track which are ``protected'' by well defined energy barriers (``pinning sites''). As a consequence, the positional stability of AFM skyrmions is expected to be greater than for their FM counterpart.

 \begin{figure}[htb]
 \centering
\includegraphics[width=0.99\columnwidth]{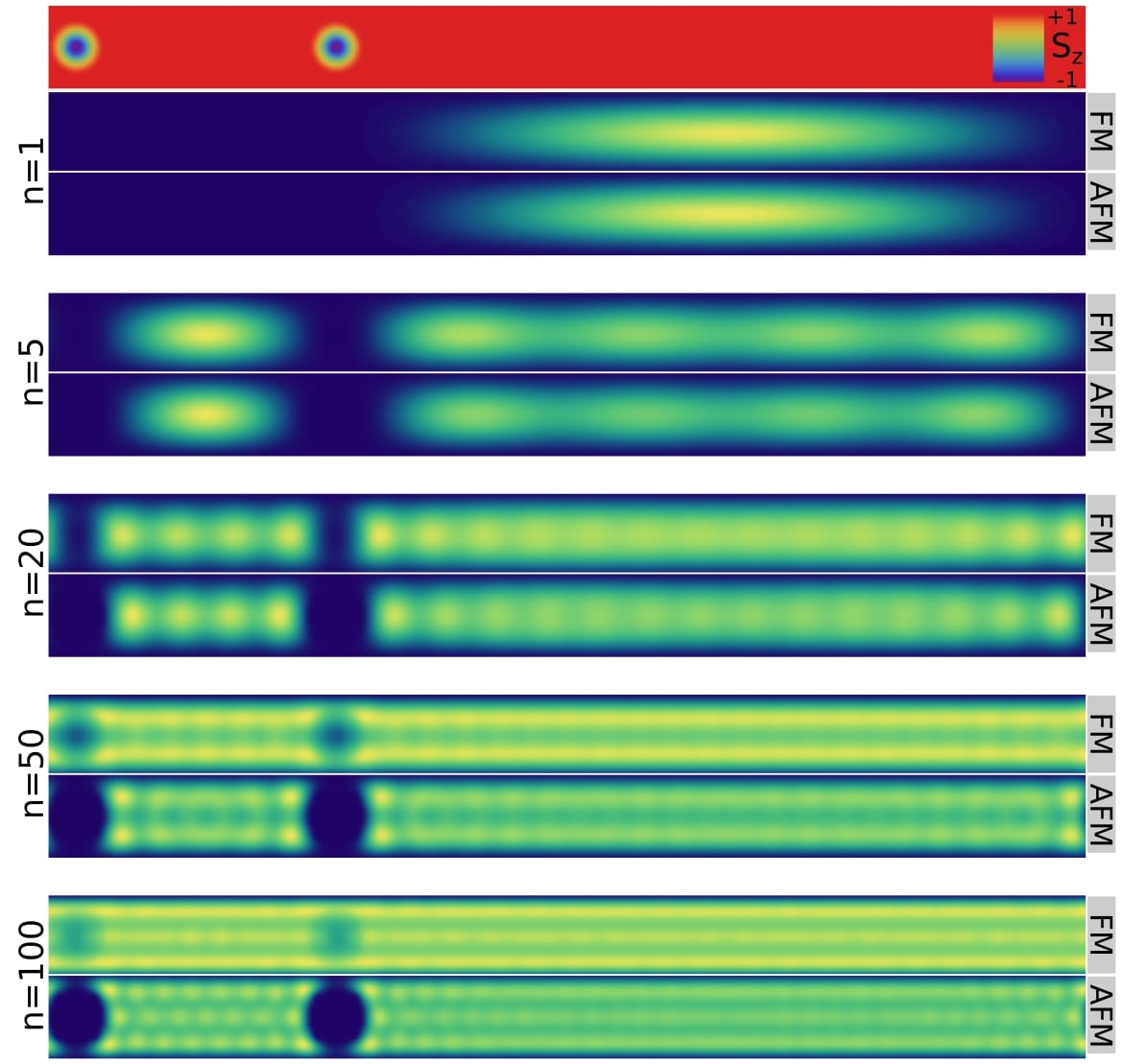}
\caption{(Color online) Electronic occupation in real space at different fillings and for ferromagnetic (FM) and antiferromagnetic (AFM) skyrmions configurations. For small fillings, both types of skyrmions induce a similar electronic distribution (for example for $n=1,5, 20$). However, as filling increases, there appears a non-zero probability that electrons will penetrate into the FM skyrmions ($n=50, 100$).The situation is completely different in the case of AFM skyrmions where the local antiferromagnetic character of the texture generates energy barriers that prevent electronic penetration.} 
\label{fig:rho_i}
\end{figure}

In order to analyze the spatial localization of the electronic states, we calculate the electronic occupation $\rho_{\rv}$ with the model described in section \ref{sec:Model}, presenting the results in Fig.\,\ref{fig:rho_i}. At the top of the figure we show the spin texture consisting of a skyrmion pair (FM-FM or AFM-AFM) separated by a distance $\Delta x$. 
Below, we present the local electron density distributions at different $n$ fillings for both cases. Quite remarkably, our initial observation was that for low fillings (as an example we show $n=1,5$) both cases present very similar distributions, showing zero electronic penetration within the skyrmion region. 

However, for larger fillings ($n=20,50$ in Fig.\,\ref{fig:rho_i}) in the FM-FM case a non-zero local electron density can be seen inside the skyrmion region ($r <R$). This is in sharp contrast to the AFM-AFM case, where the charge density  inside the skyrmion vanishes.

\begin{figure}[htb]
 \centering
\includegraphics[width=0.99\columnwidth]{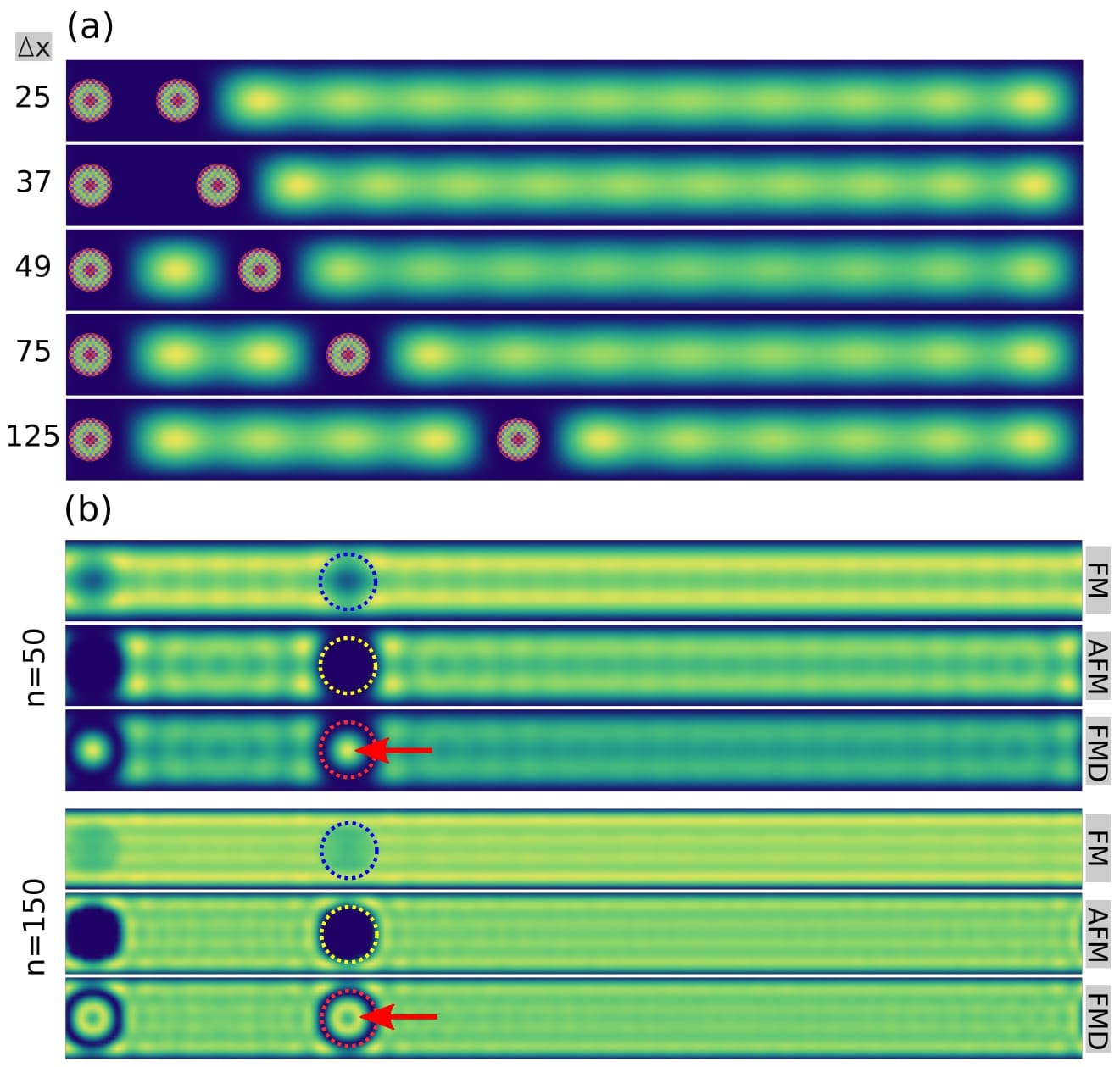}
\caption{(Color online) (a) Electronic occupation in real space for a filling $n=10$ and for different skyrmion (AF) distances $\Delta x$. (b) Comparison of the electronic occupation between  FM-FM, AFM-AFM skyrmion pairs and pairs of ferromagnetic domains FMD-FMD (with net magnetization opposite to the external field) of equal size. It can be seen that for large fillings, electronic states appear located inside the FMD's.} 
\label{fig:rho_i2}
\end{figure}

From a theoretical point of view, let's remember that the Berry phase, i.e. the quantum-mechanical phase picked up by electrons when their spin follows the orientation of the local magnetization $\Sp_\rv$, can be rewritten as an effective Aharonov-Bohm phase $a_{\rv,\rv'}$ associated with an ``emergent'' local magnetic field $B^{z}_\rv=\frac{1}{2}\Sp_{\rv}\cdot(\partial_x\Sp_{\rv}\times\partial_y\Sp_{\rv})$. In the case of FM skyrmions we expect this effective field to have a smooth dependence with the position, while this would not necessarily be the case for AFM skyrmions. This can be verified using the ans\"atze of our study to calculate  the effective fields  for both cases, obtaining $B^{z,AFM}_\rv=(-1)^{x+y}B^{z,FM}_\rv$. For the AFM case, there is a rapid oscillation of the effective field, which translates into a strong barrier potential for the electrons that prevents transmission through them. This picture is maintained for different separations between skyrmions as can be seen in Fig.\,\ref{fig:rho_i2} (panel (a)) showing the electronic distribution for the case of AFM skyrmions for different values of the distance $\Delta x$.

It should be noted that this behavior is intrinsic to AFM skyrmions. This can be appreciated much better if we consider the electronic occupation for FM-FM, AFM-AFM skyrmions and circular homogeneous ferromagnetic domains (FMD). In Fig.\,\ref{fig:rho_i2}(b) we make the comparison of the three magnetic textures for large fillings:  FM-FM skyrmions have a non-zero density inside; zero penetration into its core is observed in AFM-AFM skyrmions; while in the case of FMD-FMD localized circular states are observed inside the ferromagnetic domains.

\subsubsection{Analogy with the double-well potential problem}
\begin{figure}[htb]
\centering
\includegraphics[width=0.99\columnwidth]{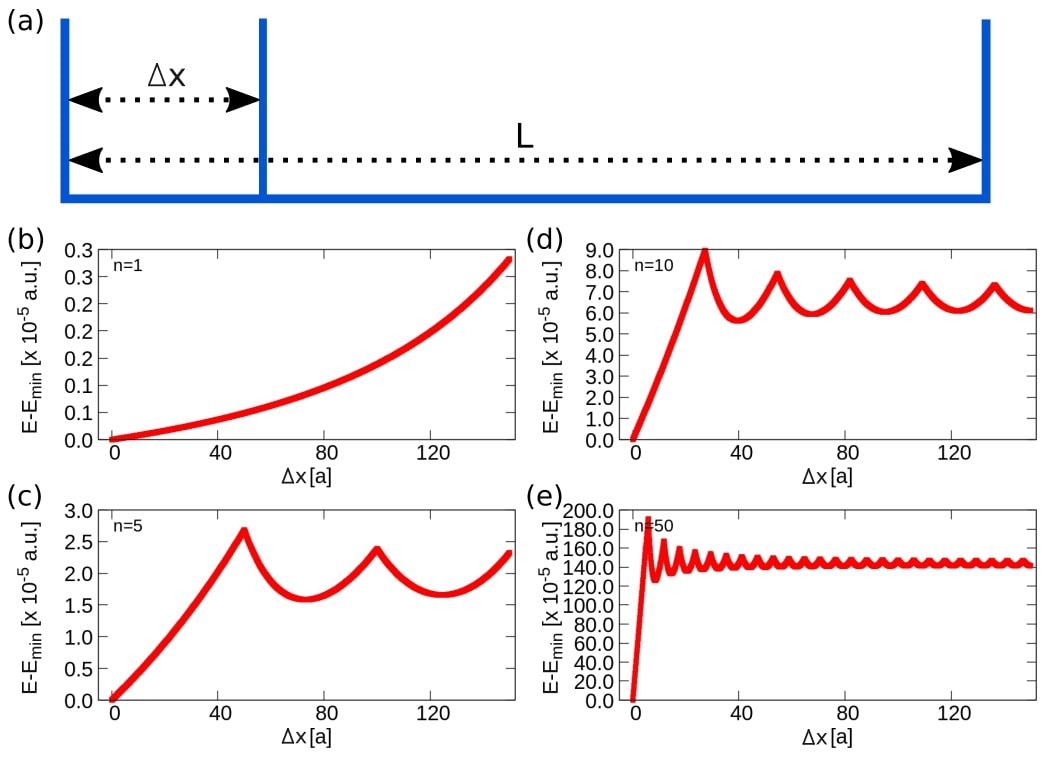}
\caption{(Color online) (a) Double potential well problem. In panels (b)-(e), system energy as a function of the distance $\Delta x$. We can see that the interaction potential is almost the same as in the case of a couple of AFM skyrmions with topological charge  $\{Q,Q\}=\{+1,+1\}$ in Fig. \ref{fig:EvsdxCurves}.} 
\label{fig:rho_i3}
\end{figure}

The exactly vanishing electronic density inside the AFM skyrmions leads us to suppose that they can be treated as impenetrable (circular) barriers for the electrons. Therefore, we may draw a parallel between the situation of itinerant electrons coupled to two skyrmions in a racetrack geometry (with skyrmion diameter comparable to the width of the racetrack) with the  quantum-mechanical textbook problem of a non-relativistic  particle confined in  an  infinite double well, where the AFM skyrmions play the role of the barriers. Let us recall that the electronic energies of non-relativistic free electrons of mass $m$ in a double well (DW) configuration (see Fig.\,\ref{fig:rho_i3}) are given by $E^{(k,l)}_{DW}(\Delta x)=\{\frac{C\,k^2}{\Delta x^2},\frac{C\,l^2}{(L-\Delta x)^2}\}$, $k,l=1,...$; $C=2\hbar^2\pi^2/2m$.  Then the energy of the system at fixed filling $n$ is 

\begin{eqnarray}
\centering
E_{DW}(\Delta x)&=&\sum_{k,l|k+l=n}E^{(k,l)}_{DW}(\Delta x).
\end{eqnarray}
In Fig. \ref{fig:rho_i3}(b-e) we present several curves of the interaction potential $E_{DW}(\Delta x)-E_{min}$. One can quickly appreciate the great similarity between the energy curves in the figures \ref{fig:rho_i3}(b-e) and the interaction potential $E(\Delta x)$ in the case of AFM skyrmions (see Fig. \ref{fig:EvsdxCurves}). The presence of a global minimum at $\Delta x=2\times R$ and the array of energetically stable positions (separations $\Delta x$) reinforces the idea that AFM skyrmions can be seen as impenetrable barriers for the itinerant electrons.

\subsection{FM-FM and AFM-AFM pairs: skyrmions-antiskyrmion and biskyrmion-biskyrmion interaction potential}
\begin{figure}[htb]
 \centering
\includegraphics[width=0.99\columnwidth]{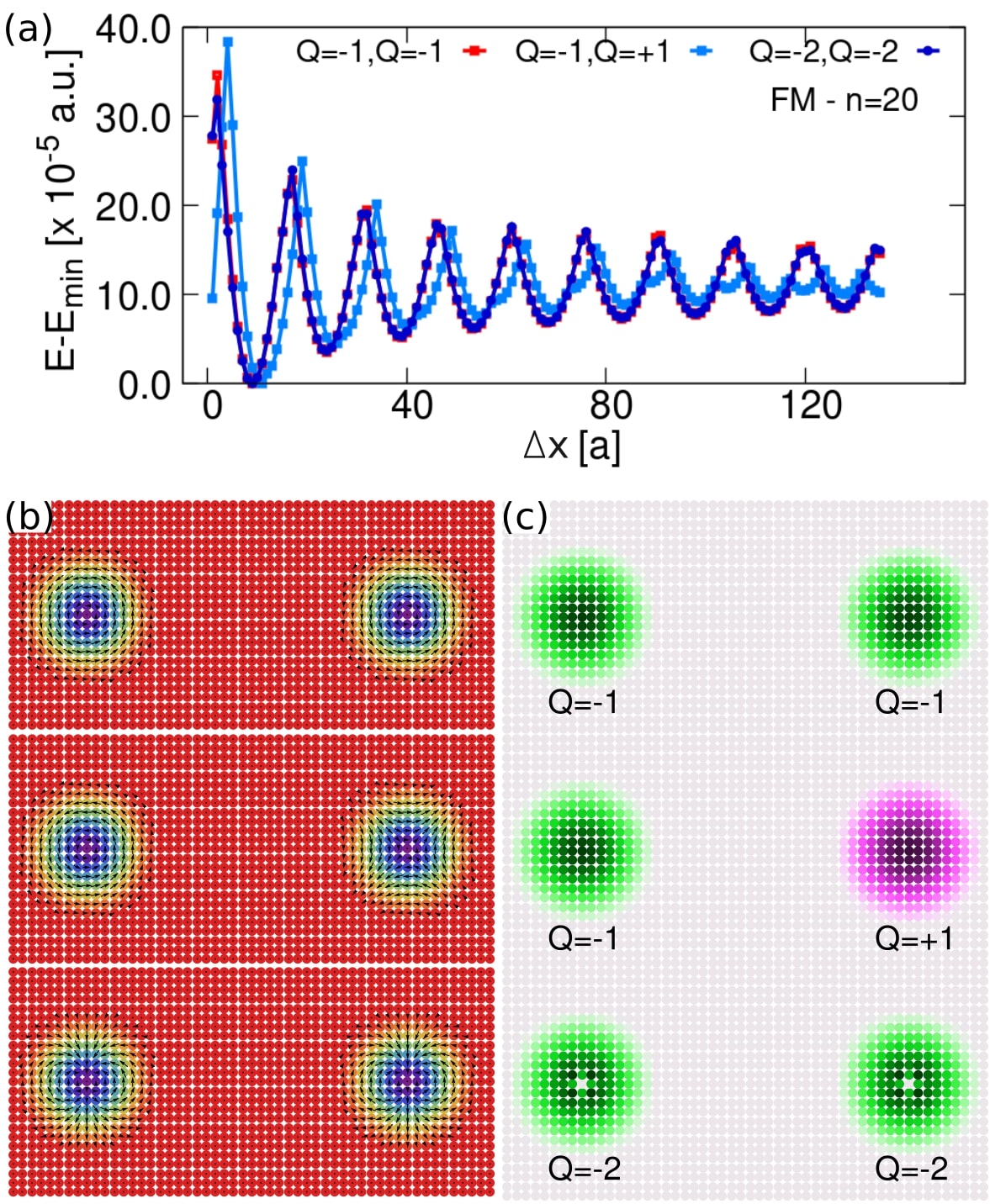}
\caption{(Color online) (a) Interaction potential $E(\Delta x)$ as a function of the horizontal distance between the FM skyrmions  with topological charge $\{Q,Q\}=\{-1,-1\}$ (red), $\{Q,Q\}=\{-1,+1\}$ (cyan) and $\{Q,Q\}=\{-2,-2\}$ (blue); (b) magnetic texture and (c) the calculated topological charge density.} 
\label{fig:rho_i4A}
\end{figure}

\begin{figure}[htb]
 \centering
\includegraphics[width=0.99\columnwidth]{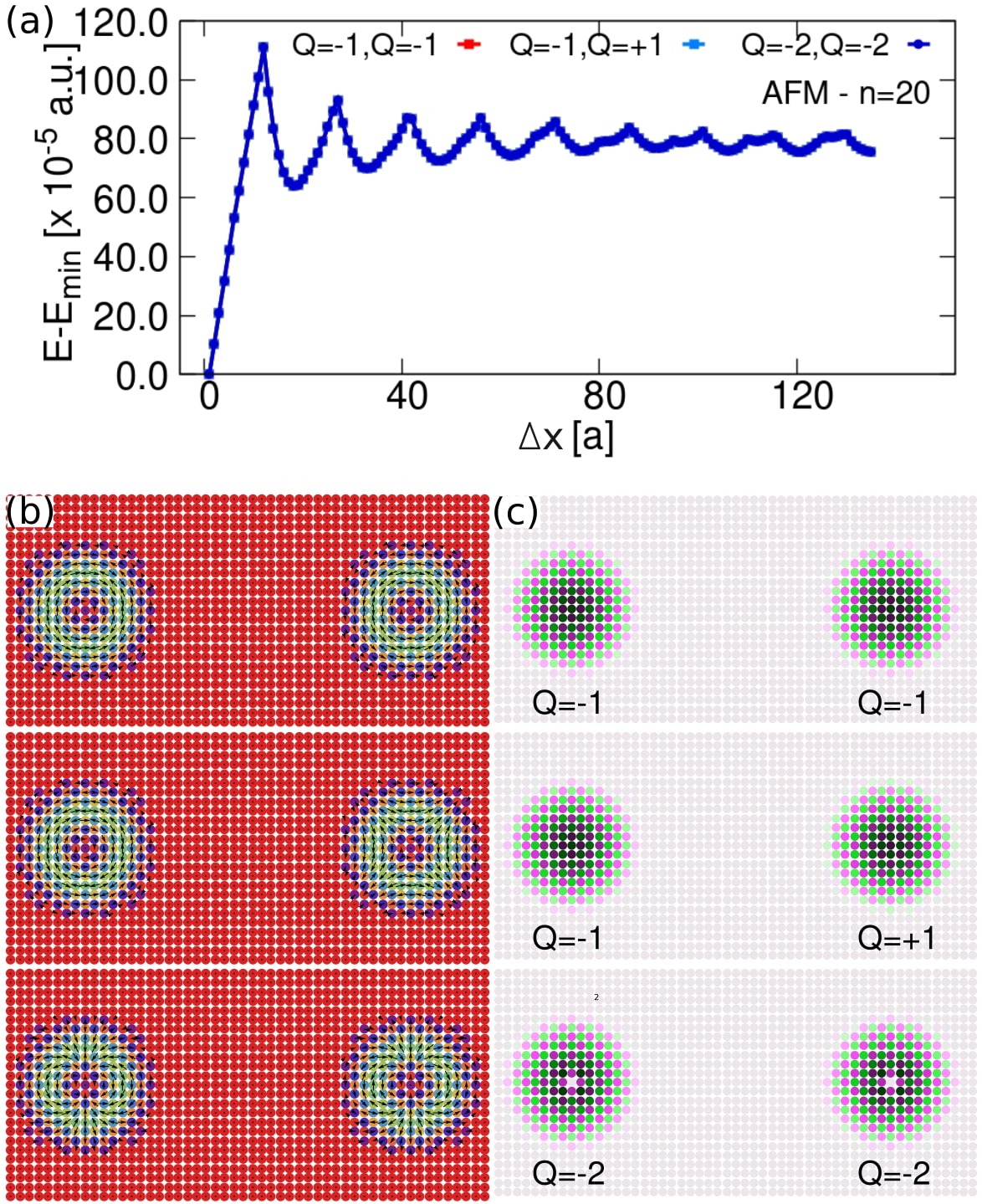}
\caption{(Color online) (a) Potential interaction $E(\Delta x)$ as a function of the horizontal distance between the AFM skyrmions with topological charge $\{Q,Q\}=\{-1,-1\}$ (red), $\{Q,Q\}=\{-1,+1\}$ (cyan) and $\{Q,Q\}=\{-2,-2\}$ (blue); (b) magnetic texture and (c) the calculated topological charge density.} 
\label{fig:rho_i4B}
\end{figure}

Up to now, we have focused on FM-FM and AFM-AFM skyrmions with topological charge $Q=-1$. As a final analysis, we study the effect of introducing skyrmions with different topological charge, i.e. antiskyrmions ($Q=+1$) and biskyrmions ($Q=-2$). In order to show this, we have explicitly computed the interaction potential $E(\Delta x)$ as a function of the horizontal distance between the FM-FM and AFM-AFM skyrmions for two specific cases: $\{Q,Q\}=\{-1,+1\}$ (skyrmion-antiskyrmion) and $\{Q,Q\}=\{-2,-2\}$ (biskyrmion-biskyrmion). First, we analyze the interaction potential $E(\Delta x)$ for FM skyrmions is displayed in Fig. \ref{fig:rho_i4A}(a). We see that the cases $\{Q,Q\}=\{-1,-1\}$ and $\{Q,Q\}=\{-2,-2\}$ exhibit almost the same behavior. The case $\{Q,Q\}=\{-1,+1\}$ presents slight differences that arise from the opposite sign of the topological charges, although it conserves the structure of local minima observed in the other cases.
In the case of AFM skyrmions (Fig. \ref{fig:rho_i4B}(a)) we obtain a perfect agreement for the three cases studied,  suggesting that in the AFM case the topological charge has a much smaller effect than in the FM case.

This great similarity between the three cases of topological charges can be understood from inspection of the local field felt by electrons: $B^{z}_{\rv}$. In Figs. \ref{fig:rho_i4A}(c) and \ref{fig:rho_i4B}(c) we present the calculated effective field associated with the configurations \ref{fig:rho_i4A}(b) and \ref{fig:rho_i4B}(b) respectively. We can see that, except for the central site, all the distributions of $B^{z}_{\rv}$ present the same structure (and opposite sign in the case $Q=+1$). This indicates that the results obtained in the case of $\{Q,Q\}=\{-1,-1\}$ are actually valid for diferent topological charges.

%

\section{Conclusions}
\label{sec:conclusions}

In recent years, magnetic skyrmions have emerged as promising candidates for devices for memory and logic applications. From the experimental point of view, the skyrmions often come in close proximity with each other, opening the natural question about how this can affect their stability, motion and spacial confinement. In this regard, we have studied the itinerant electrons-induced skyrmion-skyrmion interaction in a ferromagnetic racetrack film.  We have considered several situations: FM-FM and AFM-AFM skyrmions with three  possible  sets of topological charges: $\{Q,Q\}=\{-1,-1\}$ (skyrmion-skyrmion), $\{Q,Q\}=\{-1,+1\}$ (skyrmion-antiskyrmion) and $\{Q,Q\}=\{-2,-2\}$ (biskyrmion-biskyrmion). We found that, at low fillings, both the FM and AFM skyrmions develop a well defined array of energetically stable separations  of skyrmions along the track which are ``protected'' by well defined energy barriers. However, when filling $n$ is increased, in the case of FM skyrmions the local minima become irregular and diffuse, while in the AFM case the minima remain well defined forming a ribbed pattern as a function of $n$ for all cases. These minima in energy act as effective ``pinning sites'' along the racetrack, i.e., preferred positions for the skyrmions along the track, as long as the size of the skyrmions does not exceed the mean separation between minima. For example,  typical skyrmions have a diameter of about $6a-10a$ which corresponds to about $5$nm-$100$nm \cite{Yu2010,Nagaosa2013} .  In our study, we find that the average distance between skyrmions energy minima goes from $\sim 30a-40a$ (Fig.~(\ref{fig:EvsdxCurves}(b))) to $\sim 20a$ (Fig.~(\ref{fig:EvsdxCurves}(f))), making it possible to observe the pinning effect at low fillings. In addition, we have investigated the electronic distribution occupation. We found that at low fillings both cases (FM- and AFM-skyrmions), the electronic local density states reside outside the skyrmion region, showing zero electronic penetration within the skyrmion region. However, for larger fillings, in the FM case, a non-zero local electron density emerges inside skyrmion core. This is in sharp contrast to the AFM case even for large fillings showing that the AFM character of the skyrmions has a very large energy cost on the itinerant electrons  inside the skyrmion region.

We have confirmed that these results are valid even for other configurations of the topological charge $Q\neq -1$. In order to show this, we have explicitly computed the interaction potential $E(\Delta x)$ as a function of the horizontal distance between the FM and AFM skyrmions for two specific cases: $\{Q,Q\}=\{-1,+1\}$ and $\{Q,Q\}=\{-1,-1\}$. As a general result, we found that for FM skyrmions the cases $\{Q,Q\}=\{-1,-1\}$ and $\{Q,Q\}=\{-2,-2\}$ exhibit almost the same behavior, while the case $\{Q,Q\}=\{-1,+1\}$ presents slight differences that arise from the opposite sign of the topological charge, although it conserves the structure of local minima observed in the other cases. In the case of AFM skyrmions we observed a perfect agreement for the three cases studied,  suggesting that in the AFM case the topological charge has a much smaller effect than in the FM case. Therefore, our results support the idea that AFM skyrmions are good candidates for electronic devices. Future perspectives of this work include studying different geometries, and different skyrmions combinations such as skyrmion-antiskyrmion, FM skyrmion-AFM skyrmion, as well as different topological textures such as merons, which may also be relevant in potential technological applications.



\appendix

\section{Effective Hamiltonian Derivation}
\label{sec:Heff}
We consider a Kondo lattice model on the square lattice where the itinerant electrons are coupled with the
classical magnetic moments  by a Hund's coupling as
\bea
\mathcal{H}&=&-t\sum_{\langle \rv,\rv'\rangle,\sigma}\,(\hat{c}^{\dagger}_{\rv\sigma}\hat{c}_{\rv',\sigma}+\text{h.c.})-J_H\sum_{\rv}\Sp_{\rv}\cdot \Tp_{\rv},
\label{eq:HtJ2}
\eea

where $\hat{c}_{\rv\sigma}$ ($\hat{c}^{\dagger}_{\rv\sigma}$) is the creation (annihilation) operator at the  site $\rv$ with spin ($\sigma=\pm 1/2$), $t$  is the hopping amplitud between nearest-neighbor sites, $J_H$ is the Hund's coupling strength between the electron spin $\Tp_{\rv}=\frac{1}{2}\hat{c}^{\dagger}_{\rv,\mu}\vec{\sigma}^{\mu\nu}\hat{c}_{\rv,\nu}$ and the local magnetic moment $\Sp_\rv$.
In the strong Hund coupling limit $J_H \gg t$, the spin of the itinerant electron is fully aligned with magnetic moment $\Sp_{\rv}$. Therefore, it is trivial to observe that the electronic spectrum splits into a low- and high-energy band set \cite{tome2021topological,ohgushi2000spin}. 

In order to obtain an effective Hamiltonian describing the low energy sector we choose the quantization axis in the site $\rv$ pointing along the direction of the local magnetization, so we introduce the unitary transformation $\mathcal{U}$ between the fermionic operators, $\hat{c}^{\dagger}_{\rv}=\{\hat{c}^{\dagger}_{\rv\uparrow},\hat{c}^{\dagger}_{\rv\downarrow}\}$ and $\hat{f}^{\dagger}_{\rv}=\{\hat{f}^{\dagger}_{\rv\uparrow},\hat{f}^{\dagger}_{\rv\downarrow}\}$),   such that $\hat{c}_{\rv}=\mathcal{U}_{\rv}\cdot\hat{f}_{\rv}$ and $\mathcal{U}^{\dagger}_{\rv}\cdot(\Sp_{\rv}\cdot\vec{\sigma})\cdot\mathcal{U}_{\rv}=\sigma_z$.
The general expression of the matrix transformation is given by

\bea
\mathcal{U}^{\dagger}_\rv = {\bf m}_\rv \cdot \vec{\sigma} = \begin{pmatrix} \cos{\frac{\theta_{\rv}}{2}} & \sin{\frac{\theta_{\rv}}{2}} e^{-i \phi_{\rv}} \\ \sin{\frac{\theta_{\rv}}{2}} e^{i \phi_{\rv}} & -\cos{\frac{\theta_{\rv}}{2}} \end{pmatrix},\nonumber
\eea

where vector ${\bf m}_{\rv} = \{\sin{\frac{\theta_{\rv}}{2}} \cos{\phi_{\rv}},  \sin{\frac{\theta_{\rv}}{2}} \sin{\phi_{\rv}} , \cos{\frac{\theta_{\rv}}{2}} \}$ is defined from the local magnetic moment $\Sp_{\rv}=(\cos\phi_{\rv}  \sin\theta_{\rv}, \sin\phi_{\rv}  \sin\theta_{\rv}, \cos\theta_{\rv})$. With all this,  the transformed Hamiltonian reads

\bea
\mathcal{H}&=&-t \sum_{\rv\rv'}\hat{f}^{\dagger}_{\rv} \mathcal{U}^{\dagger}_\rv \mathcal{U}_{\rv'} \hat{f}_{\rv'} - J_H \sum_{\rv} \hat{f}^{\dagger}_{\rv} \sigma^{(z)} \hat{f}_{\rv},
\eea

where

\begin{eqnarray}
\nonumber
\mathcal{U}^{\dagger}_{\rv} \mathcal{U}_{\rv'}=
\begin{pmatrix} C_{12}  & C_{12} \\ C_{12}^{*} &C_{11}^{*}
\end{pmatrix},
\end{eqnarray}

and $C_{11}=\cos{\frac{\theta_{\rv}}{2}} \cos{\frac{\theta_{\rv'}}{2}} + \sin{\frac{\theta_{\rv}}{2}} \sin{\frac{\theta_{\rv'}}{2}} e^{-i (\phi_{\rv} - \phi_{\rv'})}$ and $C_{12}=\cos{\frac{\theta_{\rv}}{2}} \sin{\frac{\theta_{\rv'}}{2}} e^{-i \phi_{\rv'}} - \cos{\frac{\theta_{\rv'}}{2}} \sin{\frac{\theta_{\rv}}{2}} e^{-i \phi_{\rv}}$.

In the strong coupling regime $J_H \gg t$, the low-energy sector can be described by effective spinless fermions: $\{\hat{d}_{\rv},\hat{d}^{\dagger}_{\rv}\}$. As a result, the effective Hamiltonian of the system is
\bea
\nonumber
\mathcal{H}_{eff} &=&-t\sum_{\langle\rv\rv'\rangle}\cos{\left(\frac{\theta_{\rv\rv'}}{2}\right)} e^{ i a_{\rv\rv'}} \hat{d}^{\dagger}_{\rv} \hat{d}_{\rv'},
\eea
where $\hat{d}^{\dagger}_{\rv}$ ($\hat{d}_{\rv}$) corresponds to the up component of $\hat{f}^{\dagger}_{\rv}$ ($\hat{f}_{\rv}$), where $\cos\theta_{\rv\rv'}=\Sp_{\rv}\cdot\Sp_{\rv'}$ and $a_{\rv\rv'}$ the phase accumulated by the hopping electron,
\bea
\nonumber
a_{\rv\rv'} = \arctan{ \left( \frac{ -\sin{ (\phi_{\rv}-\phi_{\rv'}) } }{ \cos{ (\phi_{\rv}-\phi_{\rv'}) } + \cot{ \left( \frac{\theta_{\rv}}{2} \right)} \cot{ \left( \frac{\theta_{\rv'}}{2}\right)}} \right) }.
\eea
%



\vspace{1cm}


\begin{thebibliography}{99}
%
\bibitem{muhlbauer2009skyrmion} S. M\"uhlbauer, B. Binz, F. Jonietz, C. Pfleiderer, A. Rosch, A. Neubauer, R. Georgii, and P. B\"oni, Science {\bf 323}, 915 (2009).
%
\bibitem{munzer2010skyrmion} W. M\"unzer, A. Neubauer, T. Adams, S. M\"uhlbauer, C. Franz, F. Jonietz, R. Georgii, P. B\"oni, B. Pedersen, M. Schmidt, {\it et al}., Phys. Rev. B {\bf 81}, 041203 (2010).
%
\bibitem{yu2010real}  X. Yu, Y. Onose, N. Kanazawa, J. Park, J. Han, Y. Matsui, N. Nagaosa, and Y. Tokura, Nature {\bf 465}, 901 (2010).
%
\bibitem{yu2011near}  Yu, X., Kanazawa, N., Onose, Y. {\it et al}. Nat. Mater. {\bf 10}, 106-109 (2011).
%
\bibitem{heinze2011spon}  Heinze, S., von Bergmann, K., Menzel, M. {\it et al}. Nat. Phys. {\bf 7}, 713-718 (2011).
%
\bibitem{seki2012observation}  S. Seki, X. Yu, S. Ishiwata, and Y. Tokura, Science 336, {\bf 198} (2012).
%
\bibitem{butenko2010stab}  A. B. Butenko, A. A.  Leonov, U. K. R{\"o}{\ss}ler, and A. N. Bogdanov, Phys. Rev. B {\bf 82}, 052403 (2010).
%
\bibitem{fert2013skyrmions}  Fert, A., Cros, V. and Sampaio, J. Skyrmions on the track. Nat. Nanotech. {\bf 8}, 152-156 (2013).
%
\bibitem{zhang2015magnetic}  X. Zhang, Y. Zhou, M. Ezawa, G. Zhao, and W. Zhao, Sci. Rep. {\bf 5}, 1 (2015).
%
\bibitem{zhang2020skyrmion}  X. Zhang, Y. Zhou, K. M. Song, T.-E. Park, J. Xia, M. Ezawa, X. Liu, W. Zhao, G. Zhao, and S. Woo, J Phys Condens Matter. {\bf 32}, 143001 (2020).
%
\bibitem{GUAN2022168852} S.H. Guan and Y. Yang and Z. Jin and T.T. Liu and Y. Liu and Z.P. Hou and D.Y. Chen and Z. Fan and M. Zeng and X.B. Lu and X.S. Gao and M.H. Qin and J.-M. Liu, J. Magn. Magn. Mater: {\bf 546}, 168852 (2022).
%
\bibitem{Santos2021} V. L. Carvalho-Santos,  M. A. Castro,  D. Salazar-Aravena,  D. Laroze,  R. M. Corona,  S. Allende, and  D. Altbir, Appl. Phys. Lett. {\bf 118}, 172407 (2021).
%
\bibitem{Eschenfelder1980} Eschenfelder, A. H. Magnetic bubble technology (Springer-Verlag Berlin Heidelberg New York, 1980).

\bibitem{rosales2015three}  H. D. Rosales, D. C. Cabra, and P. Pujol, Phys. Rev. B {\bf 92}, 214439 (2015). 
%
\bibitem{barker2016static}  J. Barker and O. A. Tretiakov, Phys. Rev. Lett. {\bf 116}, 147203 (2016).
%
\bibitem{osorio2017composite}  S. A. Osorio, H. D. Rosales, M. B. Sturla, and D. C. Cabra, Phys. Rev. B {\bf 96}, 024404 (2017).
%
\bibitem{osorio2019skyrmions}  S. A. Osorio, M. B. Sturla, H. D. Rosales, and D. C. Cabra, Phys. Rev. B {\bf 100}, 220404(R) (2019).
%
\bibitem{villalba2019field}  M. E. Villalba, F. A. G\'omez Albarrac\'in, H. D. Rosales, and D. C. Cabra, Phys. Rev. B {\bf 100}, 245106 (2019).
%
\bibitem{ShangGao2020Nat}  S. Gao, H. D. Rosales, F. A. G\'omez Albarrac\'in, V. Tsurkan, G. Kaur, T. Fennell, P. Steffens, M. Boehm, P. {\v{C}}erm{\'a}k, A. Schneidewind, E. Ressouche, D. C. Cabra, C. R\"uegg, and O. Zaharko, Nature 586, 37-41 (2020).
%
\bibitem{Rosales2022a}H. D. Rosales, F. A. G\'omez Albarrac\'in, K. Guratinder, V. Tsurkan, L. Prodan, E. Ressouche, and O. Zaharko, Phys. Rev. B {\bf 105}, 224402 (2022).
%
\bibitem{legrand2020room}  W. Legrand, D. Maccariello, F. Ajejas, S. Collin, A. Vecchiola, K. Bouzehouane, N. Reyren, V. Cros, and A. Fert, Nat. Mater. {\bf 19}, 34 (2020).
%
\bibitem{liu2021neel}  Z. Liu and H. Yang, Physica E: Low-dimensional Systems and Nanostructures {\bf 135}, 114978 (2022).
%
\bibitem{mukherjee2021af}  A. Mukherjee, D. S. Kathyat, and S. Kumar, Sci. Rep. {\bf 11}, 1 (2021).
%
\bibitem{nayak2017magnetic}  A. K. Nayak, V. Kumar, T. Ma, P. Werner, E. Pippel, R. Sahoo, F. Damay, U. K. R{\"o}{\ss}ler, C. Felser, and S. S. Parkin, Nature {\bf 548}, 561 (2017). 
%
\bibitem{yu2014bisk}  Yu, X., Tokunaga, Y., Kaneko, Y. {\it et al}. Nat. Commun. {\bf 5}, 3198 (2014).
%
\bibitem{zhou2019magnetic}  Y. Zhou, National Science Review {\bf 6}, 210 (2019). 
%
\bibitem{back2020road}  C. Back, V. Cros, H. Ebert, K. Everschor-Sitte, A. Fert, M. Garst, T. Ma, S. Mankovsky, T. Monchesky, M. Mostovoy, {\it et al}., Journal of Physics D: Applied Physics {\bf 53}, 363001 (2020).
%
\bibitem{gobel2020beyond}  B. G\"obel, I. Mertig, and O. A. Tretiakov, Physics Reports, {\bf 895}, 1-28 (2020).
%
\bibitem{jungwirth2016antiferromagnetic}  T. Jungwirth, X. Marti, P. Wadley, and J. Wunderlich, Nat. Nanotech. {\bf 11}, 231 (2016).
%
\bibitem{rosales2019frustrated}  H. D. Rosales, F. A. G\'omez Albarrac\'in, and P. Pujol, Phys. Rev. B {\bf 99}, 035163 (2019).
%
\bibitem{Djavid2020}  Nima Djavid and Roger K. Lake, Phys. Rev. B {\bf 102}, 024419 (2020).
%
\bibitem{tome2021topological}  M. Tom\'e and H. D. Rosales, Phys. Rev. B {\bf 103}, L020403 (2021).
%
\bibitem{Gobel2017}B\"orge G\"obel, Alexander Mook, J\"urgen Henk, and Ingrid Mertig, Phys. Rev. B {\bf 96}, 060406(R) (2017).

%
\bibitem{Akosa2018}C. A. Akosa, O. A. Tretiakov, G. Tatara, and A. Manchon, Phys. Rev. Lett. {\bf 121}, 097204 (2018).
%
\bibitem{chen2017skyrmion}  G. Chen, Nat. Phys. {\bf 13}, 112 (2017). 
%
\bibitem{zhang2016antiferromagnetic}  X. Zhang, Y. Zhou, and M. Ezawa, Sci. Rep. {\bf 6}, 24795 (2016).
%
\bibitem{jin2016dynamics}  C. Jin, C. Song, J. Wang, and Q. Liu, Appl. Phys. Lett. {\bf 109}, 182404 (2016).
%
\bibitem{Akosa2019}  Collins Ashu Akosa, Hang Li, Gen Tatara, and Oleg A. Tretiakov, Phys. Rev. Applied {\bf 12}, 054032 (2019).
%
\bibitem{gobel2018overcoming} G\"obel, B., Mook, A., Henk, J., Mertig, I. Phys. Rev. B, {\bf 99}, 020405(2019).
%
\bibitem{Lin2013}   S. Lin, C. Reichhardt, C. D. Batista, and A.Saxena, Phys. Rev. B {\bf 87}, 214419 (2013).
%
\bibitem{Lin2016} S. Lin and S. Hayami, Phys. Rev. B {\bf 93}, 064430 (2016).
%
\bibitem{Rozsa2016}L. R\'ozsa, A. De\'ak, E. Simon, R. Yanes, L. Udvardi, L. Szunyogh, and U. Nowak, Phys. Rev. Lett. {\bf 117}, 157205 (2016).
%
\bibitem{capic2020skyrmion}  D. Capic, D. A. Garanin, and E. M. Chudnovsky, J Phys Condens Matter. {\bf 32}, 415803 (2020).
%
\bibitem{brearton2020magnetic}  R. Brearton, G. van der Laan, and T. Hesjedal, Phys. Rev. B {\bf 101}, 134422 (2020).
%
\bibitem{bezvershenko2018stabilization} Bezvershenko, A. V., Kolezhuk, A. K., and Ivanov, B. A., Phys. Rev. B,  {\bf 97}, 054408 (2018).
%
\bibitem{zhang2016magnetic} Zhang, X., Zhou, Y. and Ezawa, M. Nat. Commun., \textbf{7}, 10293 (2016).
%
\bibitem{cacilhas2018coupling} Cacilhas, R., Carvalho-Santos, V. L., Vojkovic, S., Carvalho, E. B., Pereira, A. R., Altbir, D., N\'u\~nez, \'A. S. Appl. Phys. Lett., \textbf{113}, 212406 (2018).
%
\bibitem{osorio2019stability}  S. A. Osorio, M. B. Sturla, H. D. Rosales, and D. C. Cabra, Phys. Rev. B {\bf 99}, 064439 (2019).
%
\bibitem{bogdanov1989thermodynamically}  A. N. Bogdanov and D. Yablonskii, Zh. Eksp. Teor. Fiz {\bf 95}, 178 (1989).
%
\bibitem{diaz2021majorana}  S. A. D\'iaz, J. Klinovaja, D. Loss, and S. Hoffman, Phys. Rev. B, {\bf 104(21)}, 214501.
%
\bibitem{Yu2010} Yu, X. Z. et al., Nature {\bf 465}, 901-904 (2010). 
%
\bibitem{Nagaosa2013} Nagaosa, N., Tokura, Y. ,  Nat. Nanotech. {\bf 8}, 899-911 (2013).
%
\bibitem{ohgushi2000spin}  K. Ohgushi, S. Murakami, and N. Nagaosa, Phys. Rev. B {\bf 62}, R6065 (2000).
%
\end{thebibliography}
\end{document}